# A Vehicle-portable Ultra-stable Laser for Operating on Highways

**Dongdong Jiao[1,2,5,*], Qing Li [3,*], Jing Gao[1,2], Linbo Zhang[1,2], Mengfan Wu[1,2], Qi Zang[1,2], Jianing Wang[4], Guanjun Xu[1,2,5,**], and Tao Liu[1,2,5,**]**

[1] National Time Service Center, Chinese Academy of Sciences, Xi'an 710600, China

[2] Key Laboratory of Time Reference and Application, Chinese Academy of Sciences, Xi'an 710600, China

[3] Shaanxi Aerospace Flight Vehicle Design Key Laboratory, School of Astronautics, Northwestern Polytechnical University, Xi'an, 710072, China

[4] Xi'an Shenquan Optoelectronic Technology Co., Ltd., Xi'an 710075, China

[5] Hefei National Laboratory, Hefei 230088, China

**Corresponding authors. Email: xuguanjun@ntsc.ac.cn; taoliu@ntsc.ac.cn

## Abstract

Portable ultra-stable lasers are essential for high-precision measurements. This study presents a 1550 nm vehicle-portable ultra-stable laser designed for continuous real-time operation on highways. We implement several measures to mitigate environmental impacts, including active temperature control with a standard deviation of mK/day to reduce frequency drift of the optical reference cavity, all-polarization-maintaining fiber devices to enhance the robustness of the optical path, and highly integrated electronic units to diminish thermal effects. The performance of the ultra-stable laser is evaluated through real-time beat frequency measurements with another similar ultra-stable laser over a transport distance of approximately 100 km, encompassing rural roads, national roads, urban roads, and expressways. The results indicate frequency stability of approximately $10^{-12}/(0.01s\text{-}100\text{ s})$ during transport, about $5\times10^{-14}$/s while the vehicle is stationary with the engine running, and around $3\times10^{-15}$/s with the engine off, all without active vibration isolation. This work marks the first recorded instance of a portable ultra-stable laser achieving continuous real-time operation on highways and lays a crucial foundation for non-laboratory applications, such as mobile laser communication and dynamic free-space time-frequency comparison.

Keywords: Ultra-stable laser, vehicle-portable, frequency stability, real-time, highway

## 1. Introduction

Ultra-stable lasers have significant applications in fields such as optical clocks [1,2], generation of low phase noise microwave signals [3,4], dark matter detection [5], gravitational wave detection [6,7], laser communication [8] and free-space time-frequency comparison [9]. Currently, the primary method for achieving ultra-stable lasers involves utilizing the Pound-Drever-Hall (PDH) frequency locking technique to stabilize the laser to the resonance frequency of an optical reference cavity [10,11]. This method has enabled the most advanced ultra-stable lasers to attain a linewidth of less than 10 mHz and a frequency stability on the order of $10^{-17}$ [12-15]. These ultra-stable lasers are predominantly realized using cryogenic silicon cavities [12-14] and 48 cm long ultra-low expansion glass (ULE) cavities [15].

With the advancement of scientific projects such as space optical clocks, gravitational wave detection, and gravity field measurement [16-19], the urgency for developing transportable ultra-stable lasers has significantly increased [20-27]. Currently, the best transportable ultra-stable lasers can achieve sub-Hz linewidths and frequency stability on the order of $10^{-16}$ [16,28-31]. Furthermore, some of these lasers have demonstrated the capability for long-distance transportation over thousands of km and for applications in space [16,31-34]. It is noteworthy that during these long-distance transport processes, the ultra-stable lasers remain powered off and are only activated for use upon arrival at their destination. In light of the rapid development of mobile laser communication, dynamic free-space time-frequency comparison, and mobile laser radar, there is an increasing demand for vehicle-portable ultra-stable lasers that can operate continuously and in real-time on highways [9,35-37]. This introduces new requirements for transportable ultra-stable lasers: they must not only be operational upon reaching their destination but also capable of delivering real-time, uninterrupted ultra-stable laser output during transit. To the best of our knowledge, no research has been conducted on vehicle-portable ultra-stable lasers that can function continuously and in real-time on highways. There are primarily two key technical challenges: The complexity of sound, vibration, and shock during vehicle transportation can easily lead to system lock loss and performance deterioration; The harsh temperature fluctuations within the vehicle can result in residual amplitude modulation noise (RAM) and degradation of optical power.

In this study, we investigate a 1550 nm vehicle-portable ultra-stable laser capable of continuous real-time operation on highways. This ultra-stable laser represents an enhancement of the portable ultra-stable laser developed in 2023 [33]. To improve the resistance of the portable ultra-stable laser to external temperature variations, we implement active temperature control for the optical reference cavity, achieving a temperature stability of approximately 0.6 mK. To mitigate the effects of temperature changes, sounds and transportation vibrations on laser polarization, we replace the original spatial optical path with a polarization-maintaining fiber optic path and adopt a fiber-optic polarization-maintaining fusion splicing method to minimize RAM. To reduce heat dissipation, volume, and weight of the electronics unit, we redesign the original six boards into a single board, significantly decreasing the resource requirements and impact of the electronics unit. We place two similar ultra-stable lasers in a vehicle and measure their performance through real-time comparison. These lasers are transported approximately 100 km on highways. The experimental results indicate that the frequency stability of the ultra-stable laser during transportation—across rural roads, national roads, expressways, and urban roads—is on the order of $10^{-12}$/s. In contrast, it is about $5 \times 10^{-14}$/s when the vehicle is stationary (with the engine running) and approximately $3 \times 10^{-15}$/s when the vehicle is stationary (with the engine off). Throughout the entire testing process, no active vibration isolation table is utilized. This work provides a crucial technical foundation for the continuous real-time operation of ultra-stable lasers in non-laboratory environments, such as mobile laser communication and dynamic free-space time-frequency comparison.

## 2. System design and optimization

As illustrated in Fig. 1, this portable ultra-stable laser is installed within a vehicle and comprises three primary components: a physical unit, an optical unit, and an electronics unit. The physical unit encompasses a vacuum chamber, passive thermal shielding layers, and a cubic optical reference cavity measuring 50 mm in length, which exhibits low sensitivity to vibrations [33]. All power required for the ultra-stable laser is derived from the vehicle's battery.

When the vehicle is in motion on the highway, the primary factors influencing the performance of this portable ultra-stable laser include sound, temperature fluctuations, and vibrations. The sound primarily originates from the vehicle's engine, while the internal temperature of the vehicle is predominantly affected by the engine and various electrical appliances. Vibration can be transmitted through two main pathways: one through the tires interacting with the road and the vehicle's chassis, and the other through the vehicle's frame originating from the engine. There are two main ways for vibration to be transmitted: one is through the tires on the road and the vehicle's chassis, and the other is through the vehicle's frame from the engine. To mitigate the impact of environmental factors, including temperature, sound, and vibration, on the performance of the vehicle-portable ultra-stable laser, the several measures are implemented in this work. These include active precision temperature control, a fully polarization-maintaining fiber optic path, and a highly integrated electronics unit.

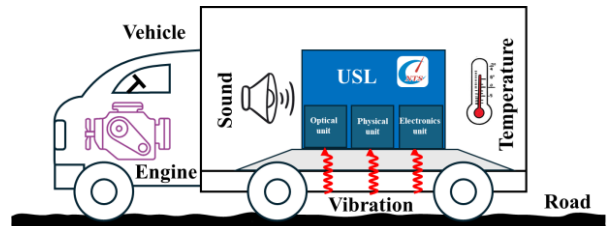

**Fig. 1.** Schematic diagram of a portable ultra-stable laser placed in a vehicle for transportation. USL denotes the portable ultra-stable laser.

### 2.1 Active precision temperature control

Compared to the laboratory environment, the temperature variation inside the vehicle is greater [38,39]. To further mitigate the impact of these temperature fluctuations on the performance of ultra-stable lasers, we implement active precise temperature control for the optical reference cavity. It is worth noting that in the initial stage, the optical reference cavity only employed passive thermal shields to suppress temperature changes [33].

Through preliminary measurements, the temperature point with zero expansion coefficient of this optical reference cavity is at about 32.5 ℃ [33], which is above room temperature and suitable for precise temperature control using unidirectional heating. As shown in Fig. 2, active precise temperature control is mainly achieved by combining heating film heating with a temperature controller. Using screws and adhesive-free glue, we tightly attach the polyimide heating film to the first layer of thermal shield, thereby providing the required temperature point for the optical reference cavity. There are a total of three thermistors (10k B3950). The first one is located on the support frame that fixes the optical reference cavity, used to monitor the temperature of the optical reference cavity. The remaining two thermistors are located on the first layer of thermal shield, one of which is used for temperature feedback, and the other is used for monitoring the heating function of the heating film. Ultimately, temperature variations in the mK range of the optical reference cavity are achieved through the temperature controller.

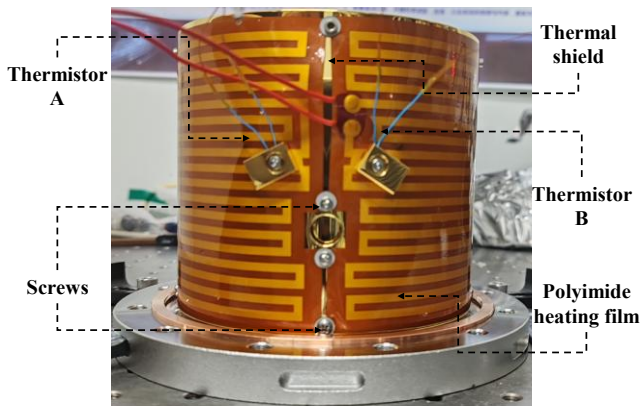

**Fig. 2.** On-site image of active precision temperature control for the cubic optical reference cavity with 50 mm length.

### 2.2 Optical unit optimization

In PDH laser frequency stabilization technology, the electro-optic modulator (EOM) is often used to achieve frequency modulation. However, it also introduces low-frequency noise generated by RAM, which prevents the signal-to-noise ratio of the frequency discrimination signal from reaching the shot noise limit, thereby reducing the frequency stability of the ultra-stable laser [40]. Changes in laser polarization are prone to cause RAM, which in turn leads to deterioration in the performance of ultra-stable lasers. Laser polarization is easily affected by temperature changes and transportation vibrations inside the vehicle [40,41]. As shown in Fig. 3, to mitigate these effects, we replace the original spatial optical path with a fiber optic path and adopt a fiber-optic polarization-maintaining fusion splicing method to minimize polarization jitter [42,43]. Through the adoption of these measures, RAM is reduced from about $3 \times 10^{-15}$/s to approximately $3 \times 10^{-16}$/s.

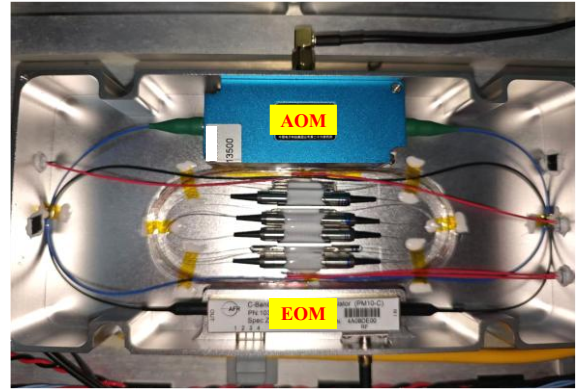

**Fig. 3.** Physical diagram of all-fiber optical path.

The optical unit of the vehicle-portable ultra-stable laser comprises the main optical path, the pre-cavity optical path, and the post-cavity optical path. The main optical path is primarily responsible for modulation and demodulation, while the pre-cavity optical path is tasked with mode matching, and the post-cavity optical path is designated for signal monitoring. To minimize the weight and volume of the optical path unit, a variety of fiber optic components—including acousto-optic modulators (AOM), EOM, and isolators—are utilized.

### 2.3 Electronics unit optimization

The primary function of the ultra-utable laser electronics unit is to provide signals and power to optical devices, while also supplying high voltage power to the vacuum system. This unit is mainly composed of six functional modules: the modem module, the PZT slow control module, the AOM fast control main module, the power supply module, the ion pump high voltage power supply module, and others. The automatic frequency locking function is the core feature of the transportable ultra-stable laser [44-46], and its hardware implementation primarily relies on the first five modules. During transportation, the ultra-stable laser is prone to losing its lock due to vibration shocks and laser frequency drift. To address these two types of lock loss, we have proposed an automatic locking algorithm that separately handles lock loss caused by vibration shocks and drift. The main control module can quickly determine the cause of lock loss: When the system is subjected to vibration impact, it uses the particle swarm optimization algorithm to obtain the optimal step value for scanning the PZT, thereby reacquiring the lock point; When the system is on the verge of losing lock due to long-term frequency drift, a pre-correction method is employed to compensate the PZT voltage in advance, restoring the lock quality before the system completely loses lock. Experimental results show that the automatic relock time is less than 1 second [46].

During transportation, vibrations and temperature changes can affect the reliability of the electronic unit and the performance of the ultra-stable laser. As shown in Fig. 4, to achieve this, we integrate six modules onto a single board, which brings four main advantages: a significant reduction in volume and weight; shortened signal transmission paths, enhancing system performance and anti-interference capability; reduced potential failure points, improving overall reliability; and simplified thermal design, avoiding local overheating issues. Through this integrated design, the volume of the electronics unit is reduced from the original 355 mm × 150 mm × 270 mm to 235 mm × 110 mm × 17 mm, a reduction of approximately 30 times.

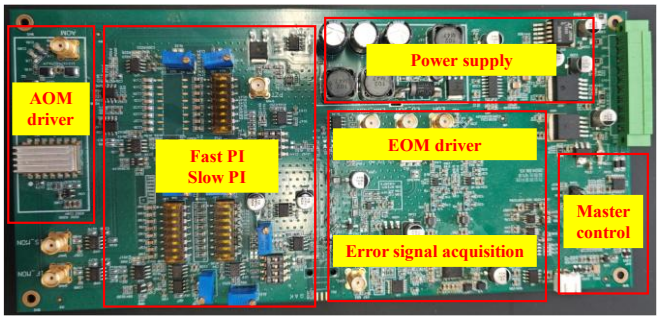

**Fig. 4.** PCB physical diagram of the electronic unit.

## 3. Experimental test design scheme

The experimental test scheme for ultra-stable lasers is shown in Fig. 5. Two ultra-stable lasers of the same structure are placed inside a vehicle, and their performance is tested through beat frequency comparison. The main testing equipment includes an oscilloscope and a frequency counter. The former is used to monitor the locking state of the ultra-stable lasers, while the latter is used to store the beat frequency signal data of the ultra-stable lasers. Each ultra-stable laser has an overall size of 8U standard chassis and weighs approximately 35 kg. An acceleration sensor is placed inside an ultra-stable laser, fixed at the upper part of the physical unit, and another acceleration sensor is fixed on the floor inside the vehicle to monitor the vibration condition inside the vehicle. It should be noted that the two ultra-stable lasers are in real-time uninterrupted beat frequency comparison.

The entire transportation test route spans approximately 100 km, encompassing four types of roads: rural roads, national roads, urban roads, and expressways. The starting point is the National Time Service Center of the Chinese Academy of Sciences (NTSC), and after traversing the aforementioned four types of roads, the route returns to the starting point, including stops at expressway toll stations and parking service areas. The duration of this transportation test is approximately 100 minutes. It should be noted that during transportation on urban roads, there are two operating conditions: braking and starting, which are encountered when facing pedestrians and traffic lights. The speed on each road is strictly in accordance with the national speed limits for motor vehicles. Within the service areas, the vehicle stops, but the engine remains running normally.

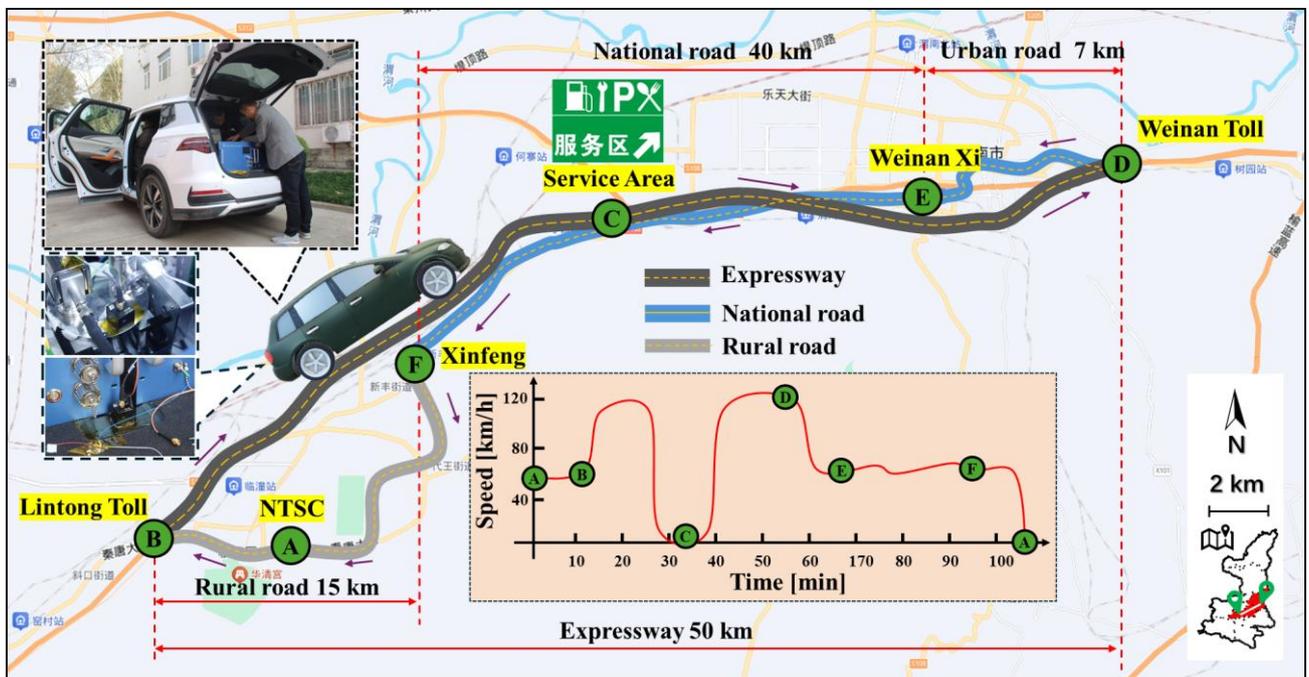

**Fig. 5.** Experimental test scheme for the vehicle-portable ultra-stable laser.

## 4. Results and discussion

The frequency variations of ultra-stable lasers within the vehicle traversing different road types are illustrated in Fig. 6. Overall, the peak-to-peak frequency variation of the ultra-stable lasers is approximately 60 kHz across rural roads, national roads, urban roads, and expressways. When the vehicle is stationary (with only the engine running), the peak-to-peak frequency variation of the ultra-stable lasers decreases to approximately 2 kHz, which is about 30 times lower than that observed during motion.

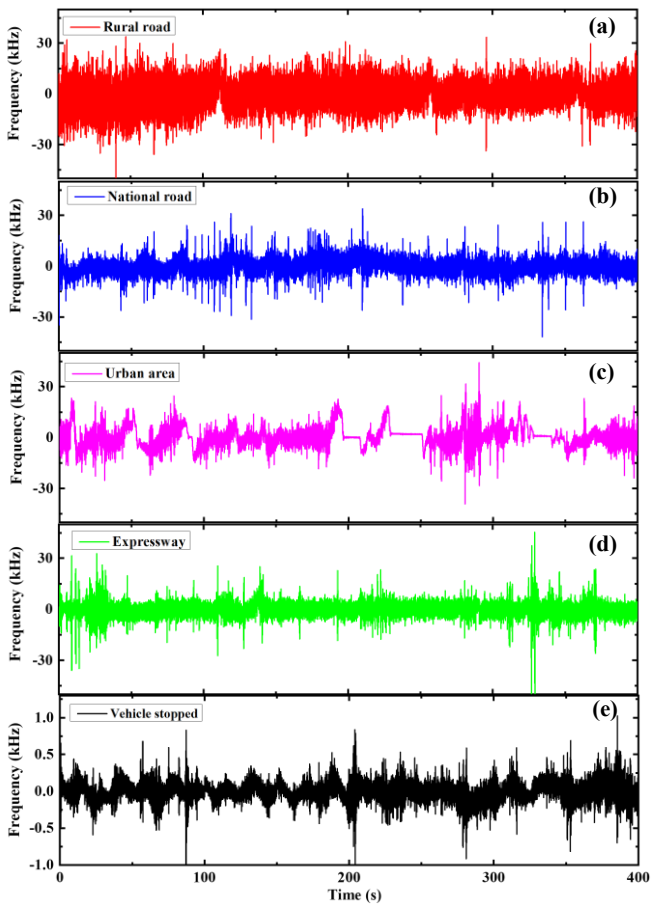

**Fig. 6.** Frequency variation of the vehicle-portable ultra-stable laser under different operating conditions.

As depicted in Fig. 6, the average frequency variation of the ultra-stable laser on rural roads is approximately 50 kHz; on urban roads, it is about 25 kHz; on national roads, it is around 20 kHz; and on expressways, the frequency variation is approximately 15 kHz. When the vehicle is stationary (with only the engine running), the average frequency variation of the ultra-stable laser is approximately 1 kHz. The primary reason for these observations may be attributed to the relative vibrations experienced on different road types, with highways exhibiting minimal vibration and rural roads experiencing significant vibration acceleration. Notably, as illustrated in Fig. 6(c), the behavior of vehicle-portable ultra-stable lasers on urban roads differs slightly from that on other road types. Specifically, during stops at traffic lights and for pedestrians, the vehicle needs to brake and start, resulting in the frequency variation of the ultra-stable lasers exhibiting abrupt changes from high values to near zero, or from near zero to high values.

The experimental results of active precision temperature control for the optical reference cavity are illustrated in Fig. 7. The black dashed-dotted line represents the temperature results within the control loop, while the blue solid line indicates the test results monitored outside the loop. The positions of the two thermistors are depicted in Fig. 2. Thermistor A is involved in temperature control within the loop, whereas Thermistor B is utilized for monitoring temperature outside the loop. The target temperature for control is set at 32.5°C, with a temperature difference within the control loop of approximately 1 mK, representing the peak-to-peak value. The average experimental temperature in the open loop is approximately 32.29°C, and the temperature difference is about 2 mK, also representing the peak-to-peak value. The standard deviation of temperature within the loop is approximately 0.2 mK, while that outside the loop is approximately 0.6 mK.

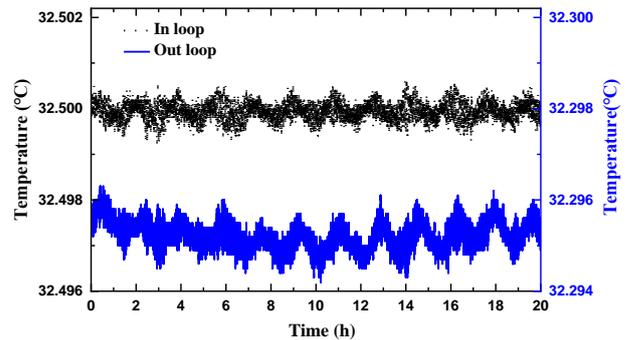

**Fig. 7.** Experimental results of active precision temperature control for the optical reference cavity.

Given the complex road conditions in urban environments, we selected this context as a typical case for analysis. The test results depicting the frequency variation of the ultra-stable laser during vehicle operation on urban roads, alongside the vibration acceleration variation of the ultra-stable laser, are illustrated in Fig. 8. The blue solid line represents the vibration acceleration variation, while the red solid line denotes the frequency variation of the ultra-stable laser. Overall, the peak-to-peak value of the frequency variation is approximately 40 kHz, and that of the vibration acceleration variation is approximately 0.5g. These test results are consistent with those reported in References [47] and [48]. Notably, between 340 seconds and 410 seconds, there were two events of braking and starting. Specifically, the braking event occurred from approximately 378 seconds to 383 seconds, while the starting event took place from approximately 399 seconds to 402 seconds. As shown in Fig. 8, during both braking and starting conditions, the response time of the frequency variation of the ultra-stable laser was faster than that of the vibration acceleration, with the former taking about 1 second and the latter about 4 seconds. Furthermore, the response of the frequency variation exhibited a more linear characteristic. These findings suggest that the ultra-stable laser is capable of measuring and rapidly responding to vibration acceleration, demonstrating superior linear characteristics.

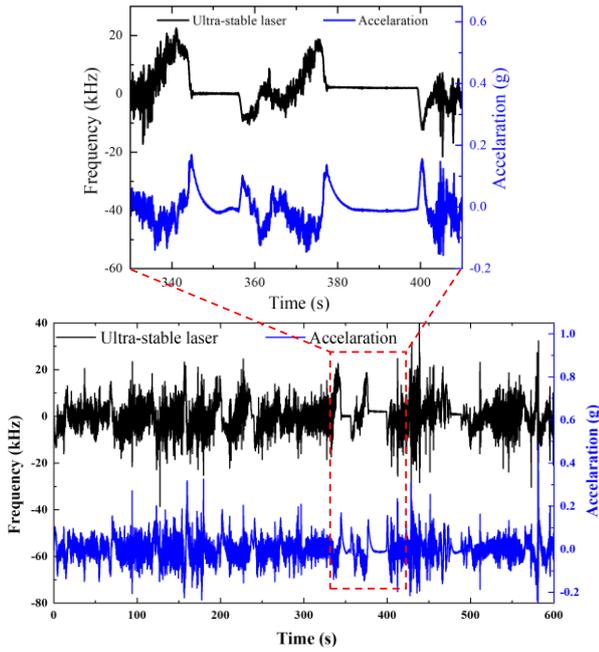

**Fig. 8.** Frequency variation and vibration acceleration of the ultra stable laser under different working conditions on the urban road.

The frequency stability of the vehicle-portable ultra-stable laser across different road types is illustrated in Fig. 9. Overall, the frequency stability of ultra-stable lasers on rural roads, national roads, expressways, and urban roads is generally on the order of magnitude of $10^{-12}$ within the time interval of 0.01 to 100 seconds, with the exception of rural roads, where the stability is on the order of magnitude of $10^{-11}$ within the time interval of 0.01 to 0.2 seconds. This discrepancy is primarily attributed to the relatively higher internal vibrations experienced by vehicles on rural roads. Within the time interval of 0.3 to 20 seconds, the frequency stability of the vehicle-portable ultra-stable laser on rural roads and national roads is essentially comparable, as the road conditions for these two types are largely similar. In this same time interval, the frequency stability of ultra-stable lasers on expressways is superior to that on national roads and urban roads, with the urban roads exhibiting the poorest stability. This is likely due to the increased presence of traffic lights on urban roads, resulting in more frequent braking and acceleration conditions.

When the vehicle is stationary (with the engine running), the frequency stability of the ultra-stable laser inside the vehicle is approximately on the order of magnitude of $10^{-14}$ within the time interval of 0.3 to 100 seconds, which is about one order of magnitude worse than that of the ultra-stable laser when the vehicle is stopped (with the engine off). However, the frequency stability on rural roads, national roads, expressways, and urban roads is approximately two to three orders of magnitude better than that of free-running laser sources [49].

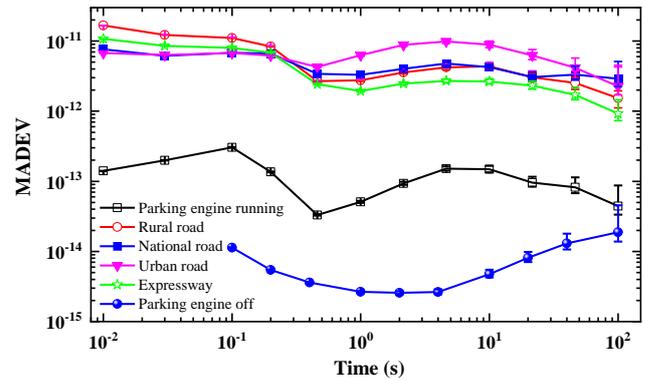

**Fig. 9.** Frequency stability of the vechile-portable ultra-stable laser under different operating conditions.

The test results indicate that this vehicle-portable ultra-stable laser exhibits strong adaptability to environmental conditions, maintaining a stable locking state across various road segments. During transportation, its frequency stability is approximately two to three orders of magnitude superior to that of a free-running laser source, yet still about three orders of magnitude inferior to its stability when stationary, primarily due to vibrations. If vibrations can be effectively mitigated, we have reason to believe that the stability of this vehicle-portable ultra-stable laser can be further enhanced. There are primarily two approaches to achieving this: the first is to improve the vibration sensitivity of the optical reference cavity, although this poses challenges under wide-ranging vibration conditions; the second is to effectively suppress the vehicle vibrations, which is a highly demanding technological endeavor. Compared to laboratory conditions, the frequency of vibrations is more complex, and the magnitude of acceleration is greater, necessitating a superior vibration isolation design [50,51]. Our subsequent work will focus on the vibration isolation design of this ultra-stable laser.

## 5. Conclusions

This study successfully develops a 1550 nm vehicle-portable ultra-stable laser designed for continuous real-time highway operation, effectively addressing the limitations of non-operable transportable lasers during transit. Three core optimizations are implemented: active precision temperature control stabilizes the optical reference cavity to approximately 0.6 mK; fiber optic paths mitigate polarization jitter caused by vibration and temperature fluctuations; and integrated electronics unit reduce both size and weight while minimizing thermal impact. Extensive 100 km multi-road tests validate the robust performance of the system: the transport-related frequency stability (approximately $10^{-12}/(0.01s\text{-}100\ s)$) is 2-3 orders of magnitude better than that of free-running lasers, with stability measurements of approximately $5\times10^{-14}/s$ when the vehicle is stationary with the engine running, and approximately $3\times10^{-15}/s$ when the engine is off, meeting near-laboratory standards without the need for active vibration isolation. This innovative technology paves the way for practical non-laboratory applications, such as mobile laser communication and dynamic free-space time-frequency comparison, thereby advancing the field of transportable ultra-stable lasers.

## Acknowledgements

This project was supported by the National Natural Science Foundation of China (Grant No. 12303076 and 12403078), the Key Research and Development Program of Shaanxi Province (Grant No. 2024GX-ZDCYL-01-28 and 2025CY-YBXM-0060), the Sanqin Talents' Special Support Program (Grant No. 09R0557A00), the Shaanxi Provincial Outstanding Young Scientist Fund Project (Grant No. 2025JC-JCQN-082), the Shaanxi Province Qin Chuangyuan "Scientist + Engineer" Team Construction Project (Grant No. 2025QCY-KXJ-162), the Open Fund of Hubei Luojia Laboratory (Grant No. 230100030), and the Science and Technology Innovation 2030 — Major Project on "Quantum Communication and Quantum Computers"(Grant No. 2021ZD0300900).

**# Dongdong Jiao and Qing Li contributed equally to this work.**